\documentstyle[11pt]{article}\textheight 230mm\textwidth 150mm
            \newcommand{\be}{\begin{eqnarray}}
            \newcommand{\ee}{\end{eqnarray}}
           \newcommand{\eel}[1]{\label{#1}\end{eqnarray}}
\newcommand{\e}[1]{\label{e:#1}\end{eqnarray}}
     \newcommand{\eg}{{\em e.g.\ }}
            \newcommand{\ie}{{\em i.e.\ }}
            \newcommand{\ga}{{\gamma}}
 
            \newcommand{\la}{{\lambda}}
            
            \newcommand{\del}{{\delta}}
 \newcommand{\om}{{\omega}}

\newcommand{\bx}{\bar{x}}
\newcommand{\bp}{\bar{p}}

           \newcommand{\ra}{{\rightarrow}}

 \newcommand{\tpe}{\tilde{p}}
\newcommand{\tx}{\tilde{x}}

 \newcommand{\hp}{\hat{p}}
\newcommand{\hx}{\hat{x}}

\newcommand{\cM}{{\cal M}}

\newcommand{\bfx}{{\bf x}}
\newcommand{\bfp}{{\bf p}}

            \newcommand{\beq}{\begin{quote}}
            \newcommand{\eq}{\end{quote}}
            
            \newcommand{\al}{\alpha}
            \newcommand{\ben}{\begin{enumerate}}
            \newcommand{\een}{\end{enumerate}}
            \newcommand{\bit}{\begin{itemize}}
            \newcommand{\ei}{\end{itemize}}
    	\newcommand{\nn}{\nonumber}
            \newcommand{\r}[1]{(\ref{e:#1})}
            \newcommand{\edfl}[1]{\Label{#1}\end{df}}

\newcommand{\dif}{{\partial}}
\newcommand{\half}{\frac{1}{2}}

\begin{document}
\begin{titlepage}

\vspace*{5 mm}

\vspace*{20mm}
\begin{center}{\LARGE\bf
Extended observables}\end{center}
\begin{center}{\LARGE\bf  in Hamiltonian theories with
constraints.}\end{center}
\vspace*{3 mm}
\begin{center}
\vspace*{3 mm}

\begin{center}Simon L. Lyakhovich\footnote{On leave of absence from
Department of Theoretical Physics, Tomsk State University, Tomsk,
Russia\\E-mail:
sll@phys.tsu.ru.} and Robert
Marnelius\footnote{E-mail: tferm@fy.chalmers.se}
 \\ \vspace*{7 mm} {\sl
Institute of Theoretical Physics\\ Chalmers University of Technology\\
G\"{o}teborg University\\
S-412 96  G\"{o}teborg, Sweden}\end{center}
\vspace*{25 mm}
\begin{abstract}
In a classical Hamiltonian theory with second class constraints
the phase space functions  on the constraint surface are observables.
We give general formulas for extended observables,
which are expressions representing the observables
in the enveloping unconstrained phase space. These
expressions satisfy in the unconstrained phase space a Poisson algebra of
the same form as the Dirac bracket algebra of the observables on the
constraint surface. The general formulas involve new differential operators
that differentiate the Dirac bracket. Similar extended observables are also
constructed for theories with first class constraints which, however,  are
gauge dependent. For such theories one may also construct gauge invariant
extensions with similar properties. Whenever  extended observables exist the
theory is expected to  allow for a covariant quantization. A mapping
 procedure is proposed for covariant quantization of theories with second
class constraints.
\end{abstract}\end{center}\end{titlepage}

\setcounter{page}{1}
\setcounter{equation}{0}
\section{Introduction.}
In this paper we present new results at a very basic level for general classical
Hamiltonian theories with constraints. We introduce the concept of extended
observables defined in a very precise way. Roughly speaking if the original
coordinates on the constraint surface are viewed as observables then extended
observables are functions of the original coordinates defined on the
unconstrained
phase space with similar properties to the observables on the constraint
surface. For
theories with second class constraints in Dirac's classifications
\cite{Dirac} the
appropriate extended observables are defined in section 3. For theories
with first
class constraints \cite{Dirac} (general gauge theories) we find three different
possible definitions of extended observables. In section 5 they are defined
in an
analogous way to the ones in section 3. The resulting  extended observables
are gauge
dependent. However, the general consensus is that observables in gauge theories
should be gauge invariant. In section 6 we present general forms for gauge
invariant
extensions. Such gauge invariant observables are well known in the
literature (see
\eg \cite{str,pol}). Now gauge theories in a particular gauge are theories with
second class constraints. When the construction of section 3 for second class
constraints is applied to such systems we obtain one very particular gauge
invariant
extension which seems to have special importance.

Apart from new very precise definitions of extended observables we also provide
simple algorithms for their constructions. These formulas involve a new
differential
operator, $V_{\al}$, which differentiates the Dirac bracket (proved in the
appendix).
The definition of this operator together with other basic formulas for Poisson
structures of theories with second class constraints is given in section 2.

The original purpose of the present work was to develop new tools for covariant
quantization of theories with second class constraints. Although these
aspects are
not developed here we make some important remarks on quantization in
section 8. First
of all we believe the existence of extended observables to be necessary for a
covariant quantization. Notice \eg that the gauge invariant extension for
the bosonic
string in \cite{str} are after quantization the DDF operators of its covariant
quantization. From the very simple models treated in this paper in section
4 it seems
as if a covariant quantization of second class constraints may be
understood from the
more conventional splitting of the constraints into gauge generators and
gauge fixing
conditions \cite{FS,HM}. However, in section 8 we also give a simple
mapping procedure
for covariant quantization which directly makes use of the extended observables.
This method is exemplified by a particle on a sphere which in this way is
quantized
in a very simple manner. The paper is then summarized in section 9.

\setcounter{equation}{0}
\section{Poisson structures in theories with second class constraints.}
Let $x^i$, $i=1,\ldots,2n$, be bosonic coordinates in a symplectic manifold
$\cM$,
$dim \cM=2n$. Let, furthermore, there be a nondegenerate two-form $\om$ on
$\cM$:
\be
&&\om=\om_{ij}(x)dx^i\wedge dx^j,\quad \det \om_{ij}\neq0,
\e{201}
which is required to be closed ($\dif_i=\dif/\dif x^i$):
\be
&&d\om=0\:\Leftrightarrow\:\dif_i\om_{jk}(x)+cycle(i,j,k)=0.
\e{202}
Since $\om$ is nondegenerate there exists an
inverse $\om^{ij}$ in terms of which the
Poisson bracket is defined by
\be
&&\{f(x), g(x)\}=\om^{ij}(x)\dif_if(x)\dif_jg(x),
\quad\om^{ij}(x)\om_{jk}(x)=\del^j_k.
\e{203}
On $\cM$ we have the natural differential operators
\be
&&\nabla^i\equiv \{x^i, \cdot\}=\om^{ij}\dif_j
\e{2031}
known as skew gradients. They satisfy a closed algebra
\be
&&[\nabla^i, \nabla^j]=\dif_k\om^{ij}\nabla^k,
\e{2032}
and Leibniz' rule
\be
&&\nabla^i\{f(x), g(x)\}=\{\nabla^if(x), g(x)\}+\{f(x), \nabla^ig(x)\},
\e{2033}
both of which follow from the Jacobi identities
\be
&&\om^{il}\dif_l\om^{jk}+cycle(ijk)=0,
\e{2034}
which in turn follow from \r{202}. $\nabla^i$ are linearly independent and
form a
basis in the tangent space. Every vector field $A$ is spanned by $\nabla^i$, \ie
we have $A=a_i\nabla^i$. $A$ differentiates the Poisson bracket \r{203} if it
satisfies Leibniz' rule, \ie
\be
&&A\{f(x), g(x)\}=\{Af(x), g(x)\}+\{f(x), Ag(x)\}.
\e{2035}
This is the case if $a_i=\dif_i a(x)$ implying that
$A$ then is a Hamiltonian vector field \ie
 $A=\{a(x), \cdot\}$. (In invariant terms, the
one-form $a_idx^i$ is then closed.)

We turn now to the constrained Hamiltonian theory. On $\cM$ we have then a
dynamical
theory with the Hamiltonian
$H(x)$ and the constraints
$\theta^{\al}(x)=0$, $\al=1,\ldots,2m<2n$, which we require  to be of
second class  in Dirac's classification \cite{Dirac}, \ie they satisfy
\be
&&\left.{\det C^{\al\beta}}\right|_{\theta=0}\neq0, \quad
C^{\al\beta}\equiv\{\theta^{\al},
\theta^{\beta}\}.
\e{204}
These constraints determine a hypersurface $\Gamma$ in $\cM$.
Notice that
\be
&&\theta^{\prime \al}(x)=0, \quad \theta^{\prime
\al}(x)=S^{\al}_{\beta}(x)\theta^{\beta}(x),\quad\left.
\det S^{\al}_{\beta}(x)\right|_{\theta=0}\neq0
\e{2041}
determine the same constraint surface $\Gamma$. The set
of constraint variables
$\{\theta^{\al}\}$ and $\{\theta^{\prime\al}\}$ are therefore equivalent.

The two-form $\om$ in \r{201} restricted to $\Gamma$ remains  a symplectic
two-form,
\ie
$\Gamma$ is a symplectic manifold. The Poisson bracket on $\Gamma$ may be
written in
terms of the coordinates $x^i$ of the enveloping manifold $\cM$. This so called
Dirac bracket \cite{Dirac} is given by
\be
&&\{f, g\}_D\equiv \{f, g\}-\{f,
\theta^{\al}\}C_{\al\beta}\{\theta^{\beta}, g\},
\e{205}
where $C_{\al\beta}$ is the inverse of $C^{\al\beta}$ in \r{204}, \ie
$C_{\al\beta}C^{\beta\ga}=\del^\ga_\al$.
The expression  \r{205} as written  is defined on the enveloping
manifold
$\cM$. However, on $\cM$ it is  a degenerate Poisson bracket since $\{f,
\theta^{\al}\}_D=0$, \ie $\theta^{\al}$ are Casimir functions for the Dirac
bracket.
(Notice that
$g=g_{\al}\theta^{\al}$ in \r{205} yields zero on $\Gamma$ but $\{f,
g_{\al}\}_D\theta^{\al}$ on $\cM$.)  The Dirac
bracket \r{205} satisfies the Jacobi identities
\be
&&\{f, \{g, h\}_D\}_D+cycle(f,g,h)=0
\e{2051}
 both on $\Gamma$ and $\cM$. Every set of equivalent
constraints lead to the same Poisson bracket {\em on the constraint surface}
$\Gamma$. However, the Dirac bracket \r{205} for different choices of equivalent
constraints {\em may be different on} $\cM$. Our constructions in the following
will be for a fixed constraint basis. The transformation properties between
equivalent sets will be studied elsewhere.

In correspondence with \r{2031} we may introduce Dirac skew gradients defined by
\be
&&D^i\equiv \{x^i, \cdot\}_D\equiv\om_D^{ij}\dif_j,\quad \om_D^{ij}\equiv\{x^i,
x^j\}_D.
\e{206}
They are linearly dependent since they satisfy the $2m$ relations
\be
&&\dif_i\theta^{\al}D^i=0,\quad\al=1,\ldots,2m.
\e{207}
$D^i$ satisfy a closed algebra and differentiate the Dirac bracket according to
Leibniz' rule, \ie we have
\be
&&[D^i, D^j]=(\dif_k\om_D^{ij})D^k.
\e{2071}
\be
&&D^i\{f, g\}_D=\{D^if, g\}_D+\{f, D^ig\}_D
\e{208}
due to  the Jacobi identities \r{2051} for the Dirac
bracket. In addition they satisfy
\be
&&D^i\theta^{\al}=0,
\e{209}
$D^i$ may, therefore, be said to differentiate parallel to the hypersurface
$\Gamma$.
The same may be said about the Dirac vector field $A_D$ defined by
$A_D\equiv a_iD^i$. $A_D$ differentiate the Dirac bracket if
$a_i=\dif_ia+\ga_{\al}\dif_i\theta^{\al}$ for any $a$ and $\ga_{\al}$
($\ga_{\al}$
does not contribute to $A_D$). Notice that the Dirac bracket \r{205} may be
written
as
\be
&&\{f, g\}_D=\om_{ij}D^ifD^jg.
\e{210}
(In
\cite{BO} this form of the Dirac bracket was applied by Batalin and
Ogievetsky in
their attempt to construct a star product on the second class surface $\Gamma$
which, however, was successful only for a  special set of constraints.)

On $\cM$ arbitrary vector fields are spanned by $\nabla^i$ in \r{2031}.
Since there
are vector fields not spanned by $D^i$ we expect the existence of more operators
differentiating the Dirac bracket. Indeed, in addition to
$D^i$ there is  another set of  $2m$ operators, $V_{\al}$, that satisfy
Leibniz' rule
with respect to the Dirac bracket. The operators $V_{\al}$ are defined by
\be
&&V_{\al}\equiv C_{\al\beta}\{\theta^{\beta},
\cdot\}\equiv C_{\al\beta}\{\theta^{\beta},
x^i\}\dif_i=C_{\al\beta}\dif_i\theta^{\beta}\nabla^i,
\e{211}
where $C_{\al\beta}$ is the inverse of \r{204} which also involved in the Dirac
bracket
\r{205}.
$V_{\al}$ are linearly independent and satisfy the properties
\be
&&V_{\al}\theta^{\beta}=\del^{\beta}_{\al},
\e{212}
and
\be
&&V_{\al}\{f, g\}_D=\{V_{\al}f, g\}_D+\{f, V_{\al}g\}_D.
\e{213}
The proof of \r{213} is nontrivial and is given in Appendix A. As far as we
know this property has not been noticed before. Notice  also that
\be
&&V_{\al}f=0\;\Leftrightarrow\{f, \theta^{\al}\}=0.
\e{214}
The following commutation relations are straight-forward to derive
\be
&&[V_{\al}, V_{\beta}]=
\dif_kC_{\al\beta}D^k \equiv \{ C_{\al\beta} , \cdot \}_D ,\nn\\
&&[D^i, V_{\al}]=-\dif_k(\{x^i, \theta^{\beta}\}C_{\beta\al})D^k  .
\e{215}
The Dirac bracket \r{205} or equivalently \r{210} may also be written in
terms of
$V_{\al}$:
\be
&&\{f,
g\}_D=\om^{ij}\dif_if\dif_jg-C^{\al\beta}V_{\al}fV_{\beta}g=
\om_{ij}\nabla^if\nabla^jg-C^{\al\beta}V_{\al}fV_{\beta}g.
\e{217}
In fact, the skew gradient $\nabla^i$ in \r{2031} may be decomposed as follows
\be
&&\nabla^i=D^i+(\nabla^i\theta^{\al})V_{\al},\;
\Leftrightarrow\;\dif_i=\om_{ik}D^k+\dif_i\theta^{\al}V_{\al}
\e{216}
$V_{\al}$ may therefore be viewed as normal
derivatives with respect to the constraint
surface $\Gamma$.  Eq.\r{216} is then a decomposition of the
derivative in parallel and normal parts.

Any vector field $\cal D$ differentiating
the Dirac bracket is decomposed as follows:
\be
&&{\cal D}\{ f , g \}_D = \{ {\cal D} f, g \}_D +
\{ f , {\cal D} g \}_D \; \Leftrightarrow \;{\cal D} =
N^\al ( \theta ) V_\al + \{ H , \cdot \}_D
\e{diff}
The coefficients $N^\al$ of the normal projections of $\cal D$
depend on $\theta$ only, whereas the parallel part is always
reduced to the action of
the Dirac bracket with some function  $H$.
The normal vector fields of $N^\al ( \theta ) V_\al$
form a closed algebra iff the constraints $\theta^\al$ form a Poisson
subalgebra in the phase space.
The question of the structure of the differentiation of a degenerate
regular Poisson bracket has been studied in the book
by Karasev and Maslov \cite{KM}. The Dirac bracket is
 a special case of a degenerate Poisson bracket, which
admits the explicit representation \r{diff}
for its differentiations. This representation
 seems to be unknown before.

\setcounter{equation}{0}
\section{Extended Observables}
The equations $\theta^{\al}(x)=0$ may be locally solved by expressing $2m$
coordinates in
terms of the remaining $2(n-m)$ independent coordinates. These solutions,
$x^{*i}$,
belong to the hypersurface $\Gamma$
which as mentioned above is a symplectic
manifold locally spanned by $2(n-m)$  independent coordinates. $x^{*i}$
represent $x^i$ in $\cM$ on $\Gamma$.
Their Poisson brackets satisfy the
relations $\{x^{*i}, x^{*j}\}=\{x^i, x^j\}_D|_{x\ra x^*}$.
We view functions of $x^{*i}$  as {\em observables}.

By  {\em extended
observables} we mean
expressions
$\bx^i(x)$ (or functions $f (\bx^i(x) )$
which are defined on the original
symplectic manifold $\cM$ and which  on $\cM$ satisfy the properties of
$x^{*i}$ on $\Gamma$. More precisely we define extended observables
$\bx^i(x)\in\cM$ to be functions satisfying the following  three properties:
\be 1)&&\bx^i(x^*)=x^{*i} \e{303} for whatever choice of solution
$x^{*i}\in\Gamma$.  \be 2)&&\theta^{\al}(\bx^i(x))=0,\quad \al=1,\ldots,2m,
\e{301}
and
\be
3)&&\{\bx^i(x), \bx^j(x)\}=\{x^i,
x^j\}_D|_{x\ra\bx(x)},
\e{302}
where the bracket on the left-hand side is the original Poisson bracket
\r{203} on
$\cM$. Notice that $\bx^i(x)$
both represents $x^{*i}$ on $\cM$ and reduces to $x^{*i}$ on $\Gamma$.
The expression \r{303} implies
that $\bx^i(x)$ must be of the general form
\be
&&\bx^i(x)=x^i+\Delta^i(x),\quad
\Delta^i(x)\equiv\sum_{k=1}^\infty
X^i_{\al_1\cdots\al_k}(x)\theta^{\al_1}(x)\cdots\theta^{\al_k}(x),
\e{304}
where  the expansion \r{304} is understood as a formal
power series in constraints $\theta^\al$ with
the coefficient
functions $X^i_{\al_1\cdots\al_k}(x)$  determined by the conditions
\r{301} and \r{302}.  The general solutions to these
conditions are derived below.

\subsection{Solving \r{301}}

First we  show that
\r{301} always have solutions of the form \r{304}. To prove this  consider
the formal Taylor expansion
\be
&&\theta^{\al}(\bx)=\theta^{\al}(x+\Delta(x))=
\theta^{\al}(x)+\Delta^i(x)\dif_i\theta^{\al}+
\half\Delta^i(x)\Delta^j(x)\dif_i\dif_j\theta^{\al}+\ldots
\e{305}
By means of this expression condition \r{301} may be solved order by order
in  powers
of
$\theta^{\al}$. To first order we get the equation
\be
&&\theta^{\al}+\theta^{\beta}X^i_{\beta}\dif_i\theta^{\al}=0,
\e{306}
and from the properties \r{209} and \r{212} we find that the vector field
\be
&&{X}_{\beta}\equiv X^i_{\beta}\dif_i
=-V_{\beta}+f_{\beta
j}D^j
\e{307}
solves \r{306} for arbitrary functions $f_{\beta j}(x)$.
Thus, we have
\be
&&X^i_{\beta}={X}_{\beta}x^i=-\{x^i,
\theta^{\ga}\}C_{\ga\beta}+f_{\beta j}\{x^j, x^i\}_D.
\e{308}
 To second order in $\theta^{\al}$ \r{301} and \r{305} yield the equation
\be
&&\left(X^i_{\beta\ga}\dif_i\theta^{\al}+
\half
X^i_{\beta}X^j_{\ga}\dif_i\dif_j\theta^{\al}\right)\theta^{\beta}\theta^{\ga}=0.
\e{309}
Again  by means of  \r{209} and \r{212} we find  that the vector field
\be
&&{X}_{\beta\ga}\equiv X^j_{\beta\ga}\dif_j=-\half
X^n_{\beta}X^m_{\ga}\dif_n
\dif_m\theta^{\rho}V_{\rho}+f_{\beta\ga
j}D^j
\e{310}
solves \r{309} for arbitrary functions $f_{\beta\ga j}$.
It is now obvious that  \r{301}   may be solved by the ansatz \r{304} order
by order in powers of $\theta^{\al}$ by means of the Taylor expansion \r{305}.
To the nth order
the solution has the form
\be
&&X^i_{\al_1\cdots\al_n}=-\Gamma^{\rho}_{\al_1\cdots\al_n}\{x^i,
\theta^{\la}\}C_{\la\rho}+f_{\al_1\cdots\al_n j}\{x^j, x^i\}_D,
\e{313}
where
$f_{\al_1\cdots\al_n j}$ are arbitrary functions of $x^i$ which are symmetric in
$\al_k$, and where $\Gamma^{\rho}_{\al_1\cdots\al_n}$ are sums of powers of
$X^i_{\al_1\cdots\al_k}$ for $k\leq n-1$ with coefficients involving
derivatives of
$\theta^{\rho}$ up to order $n$.

The above expressions may be considerably simplified. First one may remove
the derivatives of
$\theta^{\rho}$ in
$X^i_{\al_1\cdots\al_n}$ by means  of the properties
of the first order vector field $X_{\al}$ in \r{307}. From \r{307} we have
\be
&&X_{\al}\theta^{\ga}\equiv X_{\al}^i\dif_i\theta^{\ga}=-\del^{\ga}_{\al},
\e{3101}
which implies
\be
&&\dif_kX_{\al}^i\dif_i\theta^{\ga}+X_{\al}^i\dif_k\dif_i\theta^{\ga}=0,
\e{3102}
This relation allows us now to rewrite \r{310} as follows
\be
&&X_{\al_1\al_2}={1\over
4}\left(X^n_{\al_1}\dif_nX^m_{\al_2}+
X^n_{\al_2}\dif_nX^m_{\al_1}\right)\dif_m
\theta^{\ga}V_{\ga}+f_{\al_1\al_2 k}D^k,
\e{3103}
The relation \r{216} yields then
\be
&&X_{\al_1\al_2}={1\over
4}\left(X^n_{\al_1}\dif_nX^m_{\al_2}+
X^n_{\al_2}\dif_nX^m_{\al_1}\right)\dif_m
+g_{\al_1\al_2 k}D^k,\nn\\
&&g_{\al_1\al_2 k}=f_{\al_1\al_2 k}-{1\over
4}\left(X^n_{\al_1}\dif_nX^m_{\al_2}+
X^n_{\al_2}\dif_nX^m_{\al_1}\right)\om_{mk},
\e{3104}
which is much simpler than \r{310}. The same procedure may be used also for the
higher order vector fields. The third order coefficient function is \eg
determined by
the equation
\be
&&\left(X^i_{\al_1\al_2\al_3}\dif_i\theta^{\ga}+
X^i_{\al_1}X^j_{\al_2\al_3}\dif_i\dif_j\theta^{\ga}+{1\over
6}X^i_{\al_1}X^j_{\al_2}X^k_{\al_3}\dif_i
\dif_i\dif_i\theta^{\ga}\right)\theta^{\al_1}
\theta^{\al_2}\theta^{\al_3}=0,
\e{3105}
which follows from
\r{301} and \r{305}.
By means of \r{3101}, \r{3102} and
\be
&&\dif_k\dif_j X^i_{\al}\dif_i\theta^{\ga}+\dif_j
X^i_{\al}\dif_k\dif_i\theta^{\ga}+\dif_k
X^i_{\al}\dif_j\dif_i\theta^{\ga}+\dif_k
X^i_{\al}\dif_i\dif_j\dif_k\theta^{\ga}=0,
\e{3106}
which follows from \r{3102}, \r{3105} may be reduced to
\be
&&\left(X_{\al_1\al_2\al_3}^i-X^j_{\al_1\al_2}\dif_jX^i_{\al_3}+
{1\over3}X^j_{\al_1}\dif_jX^k_{\al_2}\dif_kX^i_{\al_3}-{1\over
6}X^j_{\al_1}X^k_{\al_2}\dif_j\dif_kX^i_{\al_3}\right)\times\nn\\&&\times
\dif_i\theta^{\ga}
\theta^{\al_1}\theta^{\al_2}\theta^{\al_3}=0.
\e{3107}
After insertion of the solution \r{3104} for the second order  we find
\be
&&X_{\al_1\al_2\al_3}^i={1\over6}\left(X_{\al_1}X_{\al_2}X^i_{\al_3}\right)_
{{\rm
sym}\:\al}+\left(g_{\al_1\al_2 k}D^kX^i_{\al_3}\right)_{{\rm
sym}\:\al}+\nn\\&&+g_{\al_1\al_2\al_3 k}\{x^k, x^i\}_D,
\e{3108}
where ``sym $\al$" means symmetrization in the $\al$-indices.
$g_{\al_1\al_2\al_3 k}$
is an arbitrary symmetric  function. At an arbitrary order $n$ we have in a
similar
fashion
\be
&&X_{\al_1\cdots\al_n}^i={1\over n!}\left(X_{\al_1}\cdots
X_{\al_{n-1}}X^i_{\al_{n}}\right)_{{\rm sym}\:\al}+\cdots +g_{\al_1\cdots\al_n
k}\{x^k, x^i\}_D,
\e{3109}
where the dots indicates terms involving the functions $g_{\al_1\cdots\al_m
k}$  for $m=2,\ldots,n-1$. Now these arbitrary functions  may be absorbed by a
redefinition of $f_{\al k}$ in $X^i_\al$. In other words, we may without
restrictions set $g_{\al_1\cdots\al_m
k}=0$  for $m\geq2$, and consider the functions $f_{\al k}$ to be of the form
\be
&&f_{\al k}(x)=\sum_{n=0}^\infty f^{(n)}_{\al\beta_1\cdots\beta_n
k}(x)\theta^{\beta_1}\cdots\theta^{\beta_n}.
\e{3110}
(The same
function
$\bx^i(x)$ may be obtained for different choices of the coefficient functions
$X_{\al_1\cdots\al_n}^i$ in \r{304}.)
 This redefinition implies that the
$n^{th}$ order coefficient function \r{3109} reduces to
\be
&&X_{\al_1\cdots\al_n}^i={1\over n!}\left(X_{\al_1}\cdots
X_{\al_{n-1}}X^i_{\al_{n}}\right)_{{\rm sym}\:\al}={1\over
n!}\left(X_{\al_1}\cdots
X_{\al_{n}}\right)_{{\rm sym}\:\al}x^i,
\e{3111}
which also follows from the recurrence relation
\be
&&X_{\al_1\cdots\al_n}^i={1\over
n}\left(X_{\al_1\cdots\al_{n-1}}^j\dif_jX^i_{\al_n}\right)_{{\rm
sym}\:\al},
\e{3112}
which may be derived from \r{301} and \r{305}. Eq.\r{3111}
     implies now that possible extended observables have the following simple
form in terms of the first order vector field \r{307}
\be
&&\bx^i(x)=\sum_{n=0}^\infty{1\over
n!}\theta^{\al_1}\cdots\theta^{\al_n}\left(X_{\al_1}\cdots
X_{\al_n}\right)_{{\rm
sym}\:\al} x^i\equiv\left. e^{\xi^{\al}X_{\al}}x^i\right|_{\xi=\theta},
\e{3113}
where $\xi^{\al}$ are parameters. We have the relations
\be
&&\theta^{\al}(\bx)=\theta^{\al}(\left.e^{\xi^{\beta}X_{\beta}}x\right|_{\xi
=\theta})
=\left.e^{\xi^{\beta}X_{\beta}}\theta^{\al}(x)\right|_{\xi=\theta}=0,
\e{3114}
which are valid for arbitrary $f_{\al k}$.

Notice  that the higher order coefficients in \r{3111} have   to be
symmetrized in
the $\al$-indices since $X_{\al}$ in general do not commute. We have
\be
&&[X_{\al}, X_{\beta}]=\{K_{\al\beta}, x^i\}_D\dif_i+h_{\al\beta k}D^k,
\e{312}
where
\be
&&K_{\al\beta}=C_{\al\beta}+f_{\al i}\{x^i, x^j\}_Df_{\beta
j}+C_{\beta\ga}\{\theta^{\ga}, x^i\}f_{\al i}-C_{\al\ga}
\{\theta^{\ga}, x^i\}f_{\beta
i},\nn\\
&&h_{\al\beta k}=(f_{\al i}\{x^i, x^l\}_D-C_{\al\ga}\{\theta^{\ga},
x^l\})(\dif_lf_{\beta k}-\dif_k f_{\beta l})-\nn\\&&-(f_{\beta i}\{x^i,
x^l\}_D-C_{\beta\ga}\{\theta^{\ga},
x^l\})(\dif_lf_{\al k}-\dif_k f_{\al l}).
\e{3121}

\subsection{Solving \r{302}}

The general solution \r{3113} of the condition \r{301} represents a large class
of solutions  since the functions $f_{\al k}$ are completely unconstrained
so far.
However, when we  now require the solution \r{3113}  also to  satisfy a closed
Poisson algebra in terms of the original Poisson bracket
\r{203} then the arbitrariness in $f_{\al k}$ will be considerably reduced.
Notice that if $\bx^i(x)$ satisfy a closed Poisson
algebra on $\cM$ then this algebra must coincide with the Dirac bracket
algebra, \ie we must have the
relation
\r{302}, simply since $\{\theta^{\al}(\bx),
\bx^i\}=0$  always is true.
Now already at the zeroth order in $\theta^{\al}$ we  find a restriction from
\r{302}. We find
\be
&&\{\bx^i(x), \bx^j(x)\}|_{\theta=0}=\{x^i, x^j\}+\{x^i,
\theta^{\al}\}X^j_{\al}+X^i_{\al}\{\theta^{\al}, x^j\}+X^i_{\al}\{\theta^{\al},
\theta^{\beta}\}X^j_{\beta}=\nn\\
&&=\{x^i, x^j\}_D+f_{\al k}\{x^k, x^i\}_DC^{\al\beta}\{x^l, x^j\}_D f_{\beta l},
\e{316}
which should be $\{x^i, x^j\}_D$ according to \r{302}.

Consider now the
right-hand side of the condition \r{302}.  Taylor expansion of the Dirac bracket
yields
\be
&&\{x^i,
x^j\}_D|_{x\ra\bx(x)}=\{x^i,
x^j\}_D+X^k_{\al}\dif_k\{x^i,
x^j\}_D\theta^{\al}+\nn\\&&+X^k_{\al\beta}\dif_k\{x^i,
x^j\}_D\theta^{\al}\theta^{\beta}+\half X^m_{\al}X^n_{\beta}\dif_m\dif_n\{x^i,
x^j\}_D\theta^{\al}\theta^{\beta}+\ldots\nn\\&&=
\sum_{n=0}^\infty \stackrel{(n)}{A^{ij}}, \quad \stackrel{(n)}{A^{ij}}\equiv
A^{ij}_{\al_1\cdots\al_n}\theta^{\al_1}\cdots\theta^{\al_n}.
\e{317}
To the first four orders we have
\be
&&A^{ij}=\{x^i,
x^j\}_D,\quad  A^{ij}_{\al}=X^k_{\al}\dif_k\{x^i,
x^j\}_D,
\e{3181}
\be
&&A^{ij}_{\al_1\al_2}=X^k_{\al_1\al_2}\dif_k\{x^i,
x^j\}_D+\half X^m_{\al_1}X^n_{\al_2}\dif_m\dif_n\{x^i,
x^j\}_D,
\e{3191}
\be
&A^{ij}_{\al_1\al_2\al_3}=&X^k_{\al_1\al_2\al_3}\dif_k\{x^i,
x^j\}_D+\nn\\&&+{1\over 3}\left(
X^m_{\al_1}X^n_{\al_2\al_3}+
X^m_{\al_2}X^n_{\al_3\al_1}+X^m_{\al_3}X^n_{\al_1\al_2}\right)\dif_m\dif_n\{x^i,
x^j\}_D+\nn\\&&+{1\over
6}X^k_{\al_1}X^l_{\al_2}X^m_{\al_3}\dif_k\dif_l\dif_m\{x^i,
x^j\}_D,
\e{320}
and the $n$th-order terms are symbolically given by ($\al_i$ indices and their
symmetrization are \eg suppressed)
\be
&&A^{ij}_n=\sum_{\la_1+2\la_2+\cdots +r\la_r=n}{1\over \la_1!\cdots
\la_r!}(\stackrel{(1)}{X^{k_1}})^{\la_1}\cdots (\stackrel{(r)}{X^{k_r}})^{\la_r}
(\dif_{k_1})^{\la_1}(\dif_{k_2})^{\la_2}\cdots(\dif_{k_r})^{\la_r}\{x^i,
x^j\}_D.\nn\\
\e{321}
Insertion of the solutions \r{3111} imply now
\be
&&A^{ij}_n={1\over n!}\left(X_{\al_1}\cdots X_{\al_n}\right)_{{\rm sym}\:\al}
\{x^i, x^j\}_D,
\e{322}
which means that
\be
&&\{x^i,
x^j\}_D|_{x\ra\bx(x)}=\left.e^{\xi^{\al}X_{\al}}\{x^i,
x^j\}_{D}\right|_{\xi=\theta}.
\e{323}

The question now is under which conditions this is equal to $\{\bx^i,
\bx^j\}$. We
notice then that this definitely requires  $X_{\al}$ to differentiate the Dirac
bracket. (We have explicitly checked this for the first three orders.)
Thus, $f_{\al
k}$ must be of the form
\be
&&f_{\al
k}=\sum_{n=0}^\infty
\dif_kb_{\al\beta_1\cdots\beta_n}\theta^{\beta_1}
\cdots\theta^{\beta_n},
\e{324}
where $b_{\al\beta_1\cdots\beta_n}$ are arbitrary functions of $x^i$.
This implies
\be
&&X_{\al}=-V_{\al}+\{f_{\al}, \cdot\}_D, \quad  f_{\al}=
\sum_{n=0}^\infty b_{\al\beta_1\cdots\beta_n}\theta^{\beta_1}
\cdots\theta^{\beta_n}.
\e{325}
A further consequence of \r{324} is that \r{312} reduces to
\be
&&[X_{\al}, X_{\beta}]=\{K_{\al\beta}, \cdot \}_D \, ,
\e{326}
where
\be
&&K_{\al\beta}=C_{\al\beta}+\{f_{\al}, f_{\beta}\}_D+C_{\beta\ga}\{\theta^{\ga},
f_{\al }\}-C_{\al\ga}
\{\theta^{\ga}, f_{\beta}\}.
\e{327}
The property  \r{326} is consistent
with the fact that
$X_{\al}$ in \r{325} differentiates the Dirac bracket.
The commutator of differentiations is a differentiation,
thus it must have the form \r{diff}.
$K_{\al\beta}$ \r{327} gives the explicit
expression for the potential
$H$ in the parallel part of the derivative in \r{diff}.

For $X_{\al}$ of the form \r{325} we have
\be
&&\left.e^{\xi^{\al}X_{\al}}{\{x^i,
x^j\}_D}\right|_{\xi=\theta}=\{\left.e^{\xi^{\al}X_{\al}}{x^i}\right|_{\xi=\
theta},
\left.e^{\xi^{\al}X_{\al}}{x^i}\right|_{\xi=\theta}\}_D=\{\bx^i(x),
\bx^j(x)\}_D.
\e{3301}
From \r{217} condition \r{302} requires now
\be
&&V_{\al}\bx^i(x)C^{\al\beta}V_{\beta}\bx^j(x)=0.
\e{3311}
From the general expression \r{3113} of $\bx^i(x)$  we have
\be
&&V_{\al}\bx^i(x)=\{f_{\al},\bx^i(x)\}_D-
X_{\al}\bx^i(x),\nn\\&&
X_{\al}\bx^i(x)=\half\theta^{\beta}[X_{\al},
X_{\beta}]x^i+O(\theta^2).
\e{3312}
Hence, to the zeroth order \r{3311} requires
\be
&&\{f_{\al},x^i\}_DC^{\al\beta}\{f_{\beta},x^j\}_D=0,
\e{3313}
which is consistent with the result \r{316}. The condition \r{3311}
becomes to the first order
\be
&&\left.\left(\{f_{\al}, X_{\ga}x^i\}_D+\half\{K_{\al\ga},
x^i\}_D\right)C^{\al\beta}
\left(\{f_{\beta}, X_{\del}x^j\}_D+\half\{K_{\beta\del},
x^j\}_D\right)\right|_{{\rm
sym}\:\ga\del}=0.\nn\\
\e{318}
This condition  as well as all the higher order conditions from \r{3311}
are intricate conditions on $f_{\al}$ and $K_{\al\beta}$ in \r{327}. If we
are able to
choose $f_{\al}$ such that $K_{\al\beta}$ are expressed in terms of
$\theta^{\al}$
and constants then the $K_{\al\beta}$-dependence in these conditions will
disappear.
This is exactly the condition for $X_{\al}$ to commute in \r{326}. For commuting
$X_{\al}$ we have also
\be
&&X_{\al}\bx^i(x)=0\quad\Leftrightarrow\quad V_{\al}\bx^i=\{f_{\al},\bx^i\}_D,
\e{319}
which makes the conditions \r{3311} equivalent to
\be
&&\{f_{\al},\bx^i\}_DC^{\al\beta}\{f_{\beta},\bx^j\}_D=0,
\e{334}
which seems to be a simple condition on
$f_{\al}$. In fact, all explicit solutions considered in the following have
commuting $X_{\al}$. Maybe this is a general feature. Notice that we always have
the property
\be
&&\bx^i(\bx(x))=\bx^i(x),
\e{3341}
which trivially follows if $X_{\al}$ commute due to \r{319}.

Whether or not there exist functions $f_{\al}$ satisfying both
\r{334} and
$\{K_{\al\beta}, g(x)\}_D=0$ for arbitrary functions $g(x)\in\cM$ in all
theories with
second class constraints is unclear.
However, we expect that the class of theories for which this is possible to
be large.
(In subsection 4.2  and section 7 we give simple examples with a nontrivial
$C_{\al\beta}$ which satisfy these properties.)
One may notice that all
theories for which
$C_{\al\beta}$  is a function of only the constraints (i.e., in this case,
the constraints $\theta^\al$ constitute Poisson subalgebra in the phase
space) so that $V_{\al}$ commute are contained in this class since $f_{\al}$
then may be chosen to be zero   and we have \be
&&V_{\al}\bx^i(x)=0\;\Leftrightarrow\{\bx^i(x), \theta^{\al}\}=0.
\e{339}
In this special case, all the extended observables commute with the
constraints. If the constraints do not form a Poisson subalgebra they can, of course, not commute with
 $\bx^i(x)$.

\setcounter{equation}{0}
\section{Examples}
\subsection{A simple example}
Consider a dynamical theory defined on a phase space, $\cM$,  on which
$x^A$ and $p_A$
are globally defined canonically conjugate variables. We have the
fundamental Poisson
bracket relations
\be
&&\{x^A, p_B\}=\del^A_B,\quad \{x^A, x^B\}=\{p_A, p_B\}=0.
\e{4001}
The indices A, B are assumed to be raised (and lowered) by a constant regular
symmetric metric
$g^{AB}$ ($g_{AB}$).

On the phase space $\cM$ we have two constraints, $\theta^\al=0$, where
\be
&&\theta^1=x^Ax_A-R^2,\quad \theta^2=p_Ax^A,
\e{4002}
where $R$ is a positive constant.
These constraints satisfy
\be
C^{12}\equiv\{\theta^1, \theta^2\}=2x^Ax_A=2\theta^1+2R^2.
\e{4003}
Hence, the constraints are of second class and they form closed
Poisson algebra.
The Dirac bracket is
\be
&&\{A, B\}_D=\{A, B\}-\{A, \theta^1\}C_{12}\{\theta^2,
B\}-\{A, \theta^2\}C_{21}\{\theta^1, B\}, \e{4004} where ($x^2\equiv x^Ax_A$)
\be
&&C_{12}=-{1\over x^2}=-C_{21}.
\e{4005}
Explicitly we find
\be
&&\{x^A, x^B\}_D=0,\quad \{x^A, p_B\}_D=\del^A_B-{x^Ax_B\over x^2},\nn\\&&
\{p_A,
p_B\}_D={1\over x^2}(p_Ax_B-p_Bx_A).
\e{4006}
A general ansatz for extended observables satisfying the conditions \r{304} and
\r{301} is
\be
&&\bx^A={R\over\sqrt{x^2}}x^A,\quad \bp_A=p_A-p\cdot x{x_A\over
x^2}+M_{AB}(x,p)x^B,
\e{4007}
where $M_{AB}(x,p)$ is an arbitrary antisymmetric tensor function which
vanishes for
$\theta^{\al}=0$. The condition
\r{302}, \ie the correct Poisson bracket algebra fixes $M_{AB}(x,p)$. We get the
 solution ($p\cdot x=p_Ax^A$)
\be
&&\bx^A={R\over\sqrt{x^2}}x^A,\quad \bp_A={\sqrt{x^2}\over R}(p_A-p\cdot
x{x_A\over
x^2}).
\e{4008}
Only these expressions satisfy
\be
&&\{\bx^A, \bx^B\}=0,\quad \{\bx^A, \bp_B\}=\del^A_B-{x^Ax_B\over
x^2}=\del^A_B-{\bx^A\bx_B\over \bx^2}=\del^A_B-{\bx^A\bx_B\over
R^2},\nn\\&& \{\bp_A,
\bp_B\}={1\over \bx^2}(\bp_A\bx_B-\bp_B\bx_A)={1\over
R^2}(\bp_A\bx_B-\bp_B\bx_A).
\e{4009}
Since $\{A, C^{12}\}_D=0$ according to \r{4003}, the
$V_{\al}$-operators   commute. We have
\be
&&V_1\equiv C_{12}\{\theta^2,\cdot\}={1\over2x^2}(x^A\dif_A^x-p^A\dif_A^p),\quad
V_2\equiv C_{21}\{\theta^1, \cdot\}={x^A\over x^2}\dif_A^p,
\e{4010}
where $\dif_A^x=\dif/\dif x^A$ and $\dif_A^p=\dif/\dif p^A$. One may easily
check that $[V_1,V_2]=0$. Since we have $f_{\al}=0$ here  the extended
observables
must satisfy $V_{\al}\bx^A=V_{\al}\bp_A=0$ according to
\r{339} or
equivalently
\be
&&\{\bx^A, \theta^{\al}\}=\{\bp_A, \theta^{\al}\}=0.
\e{4011}
This condition on the general ansatz \r{4007} yields again the  expressions
\r{4008}. Thus, the expressions \r{4008} are possible to write as
\be
&&\left.\bx^A=e^{-\xi^{\al}V_{\al}}x^A\right|_{\xi=\theta},\quad
\left.\bp_A=e^{-\xi^{\al}V_{\al}}p_{A}\right|_{\xi=\theta}.
\e{4012}
According to \r{217} the properties \r{4011} imply
\be
&&\{\bx^A, \bx^B\}=\{\bx^A, \bx^B\}_D,\quad \{\bx^A, \bp_B\}=\{\bx^A,
\bp_B\}_D,\quad \{\bp_A,
\bp_B\}=\{\bp_A,
\bp_B\}_D.
\e{4013}

\subsection{A simple but nontrivial example}
Consider a $2n$-dimensional phase space ($n\geq2$), $\cM$, spanned by the
canonical
coordinates $x^{\mu}$ and $p_{\mu}$ satisfying the Poisson algebra
\be
&&\{x^{\mu}, p_{\nu}\}=\del^{\mu}_{\nu},\quad\{x^{\mu}, x^{\nu}\}=\{ p_{\mu},
p_{\nu}\}=0.
\e{401}
On $\cM$ we impose two constraints, $\theta^{\al}=0$, where
\be
&&\theta^1\equiv p^2-m^2,\quad \theta^2\equiv x^2-a^2,
\e{402}
where $a$ and $m$ are two real constants. $x^{\mu}$ and $p_{\nu}$ are
considered to
be n-dimensional Lorentz vectors and all inner products are Lorentz
products. Thus, in
\r{402}
$p^2=p_{\mu}p_{\nu}\eta^{\mu\nu}$ and $x^2=x^{\mu}x^{\nu}\eta_{\mu\nu}$ where
$\eta^{\mu\nu}(\eta_{\mu\nu})$ is a time-like Minkowski metric in n
dimensions. The
constraints
$\theta^{\al}=0$ are of second class since
\be
&&\left.{C^{12}}\right|_{\theta=0}\neq0, \quad C^{12}=\{\theta^1,
\theta^2\}=-4p\cdot
x,
\e{403}
where $p\cdot x=p_{\mu}x^{\mu}$. The Dirac bracket is given by \r{4004} with
\be
&&C_{12}={1\over 4p\cdot x}=-C_{21}.
\e{405}
Explicitly we  get
\be
&&\{x^{\mu}, p_{\nu}\}_D=\del^{\mu}_{\nu}-{p^{\mu}x_{\nu}\over p\cdot x},\quad
\{x^{\mu}, x^{\nu}\}_D=\{ p_{\mu},
p_{\nu}\}_D=0.
\e{406}

In order to construct appropriate extended observables we first construct
general
solutions of
\r{301} for the ansatz
\r{304}. There are several solutions. Three of them are given below.
\be
1)&&\bx^{\mu}=x^{\mu}-{p^{\mu}(p\cdot x)\over
p^2}\left(1-\sqrt{A_x}\right),\quad
\bp^{\mu}=p^{\mu}-{x^{\mu}(p\cdot x)\over x^2}\left(1-\sqrt{A_p}\right),
\e{407}
where
\be
&&A_x=1-{p^2(x^2-a^2)\over(p\cdot x)^2}, \quad
A_p=1-{x^2(p^2-m^2)\over(p\cdot x)^2}.
\e{408}
\be
2)&&\bx^{\mu}=\sqrt{{p^2\over m^2}}\left(x^{\mu}-{p^{\mu}(p\cdot x)\over
p^2}\left(1-\sqrt{B}\right)\right),\quad\bp^{\mu}=\sqrt{{m^2\over
p^2}}\,p^{\mu},
\e{409}
where
\be
&&B=1-{p^2x^2-m^2a^2\over(p\cdot x)^2}.
\e{410}
\be
3)&&\bx^{\mu}=\sqrt{{a^2\over x^2}}\,x^{\mu},\quad\bp^{\mu}=\sqrt{{x^2\over
a^2}}\left(p^{\mu}-{x^{\mu}(p\cdot x)\over x^2}\left(1-\sqrt{B}\right)\right),
\e{411}
All three expressions \r{407}, \r{409} and \r{411} satisfy
\be
&&\bx^2=a^2,\quad \bp^2=m^2.
\e{412}
It is straight-forward but tedious to check that the expressions \r{409}
and \r{411}
satisfy
\be
&&\{\bx^{\mu}, \bp_{\nu}\}=\del^{\mu}_{\nu}-{\bp^{\mu}\bx_{\nu}\over \bp\cdot
\bx},\quad
\{\bx^{\mu}, \bx^{\nu}\}=\{\bp_{\mu},
\bp_{\nu}\}=0,
\e{413}
which are the correct expressions required by \r{302}
due to \r{406}. However, \r{407}
does not satisfy \r{413}.
The reason why we have found more than one  correct solution will be
explained in
section 8. (There might be more than two solutions.)

We consider now  the above  solutions as expansions in the
constraint variables:
\be
&&\bx^{\mu}=x^{\mu}+\sum_{k=1}^\infty X^{\mu}_{\al_1\cdots\al_k}
\theta^{\al_1}\cdots\theta^{\al_k},\nn\\
&&\bp^{\mu}=p^{\mu}+\sum_{k=1}^\infty P^{\mu}_{\al_1\cdots\al_k}
\theta^{\al_1}\cdots\theta^{\al_k}.
\e{414}
The solution \r{407} yields to first order \eg
\be
&&X^{\mu}_1=0,\quad
X^{\mu}_2=-{p^{\mu}\over 2(p\cdot x)},\quad P^{\mu}_1=-{x^{\mu}\over(p\cdot
x)},\quad
P^{\mu}_2=0.
\e{415}
Other choices which differ from these by terms involving powers of constraint
variables are also possible and are considered to be equivalent.
Comparison with the
general expression
\r{325},
\ie
\be
&&X^{\mu}_{\beta}=-\{x^{\mu}, \theta^{\ga}\}C_{\ga\beta}+\{f_{\beta},
x^{\mu}\}_D,\nn\\&&P^{\mu}_{\beta}=-\{p^{\mu},
\theta^{\ga}\}C_{\ga\beta}+\{f_{\beta},
p^{\mu}\}_D
\e{416}
yields  $f_{\al}=0$, which means that the fundamental first order vector fields
$X_{\al}$ do not commute for \r{407}. $X_{\al}$ are  given by
\be
&&X_{1}=X_1^{\mu}\dif^x_{\mu}+P_1^{\mu}\dif^p_{\mu}, \quad
X_{2}=X_2^{\mu}\dif^x_{\mu}+P_2^{\mu}\dif^p_{\mu},
\e{417}
where  $\dif^x_{\mu}=\dif/\dif x^{\mu}$ and $\dif^p_{\mu}=\dif/\dif p^{\mu}$.
Although the condition \r{334} is satisfied it is not equivalent to
\r{3311}. In fact,
we have here
\be
&&X_1\bp^{\mu}=X_2\bx^{\mu}=0,\quad X_1\bx^{\mu}\neq0,\quad
X_2\bp^{\mu}\neq0,\nn\\
&&\Rightarrow\quad X_{\al}\bx^{\mu}C^{\al\beta}X_{\beta}\bp^{\nu}\neq0,
\e{418}
which violates \r{3311} since $X_{\al}=-V_{\al}$ here. This explains why
\r{407} does
not satisfy \r{413}.

Consider now the solution \r{409}. It yields to first order \eg
\be
&&X^{\mu}_1={x^{\mu}\over 2m^2}-{a^2\over 2p^2(p\cdot x)}p^{\mu},\quad
X^{\mu}_2=-{p^{\mu}\over 2(p\cdot x)},\quad P^{\mu}_1=-{p^{\mu}\over2m^2},\quad
P^{\mu}_2=0.
\e{419}
  Comparison with the
general expression
\r{416} yields then the possible choices for $f_{\al}$. One possible choice is
\be
&&f_1=-{(p\cdot x)\over 2m^2},\quad f_2=0.
\e{420}
This choice makes \r{334} zero and yields for \r{327}
\be
&&K_{12}=C_{12}{(m^2-p^2)\over m^2}=-K_{21}.
\e{421}
Let us now see if we  can make another choice of $f_{\al}$ such that the
first order
vector fields \r{417} commute by requiring
$K_{\al\beta}$ in \r{327} to be functions of the constraint variables
$\theta^{\al}$
only. The allowed forms of $f_{\al}$ are given in \r{325}. We may therefore
replace
$f_1$ in
\r{420} by
\be
&&f_1=-{(p\cdot x)\over 2m^2}+b_{11}(p^2-m^2), \quad f_2=0.
\e{422}
This expression makes $K_{\al\beta}$ zero for the choice
\be
&&b_{11}={(p\cdot x)\over2m^2p^2} \;\Rightarrow\;f_1=-{(p\cdot x)\over2p^2}.
\e{423}
This $f_1$ inserted into \r{416} yields the first order coefficients
\be
&&X^{\mu}_1={x^{\mu}\over 2p^2}-{x^2\over 2p^2(p\cdot x)}p^{\mu},\quad
X^{\mu}_2=-{p^{\mu}\over 2(p\cdot x)},\quad P^{\mu}_1=-{p^{\mu}\over2p^2},\quad
P^{\mu}_2=0,
\e{424}
which differ from \r{419} by a power expansion in the constraint variables.
One may
easily check that ($X_{\al}$ are the first order vector field defined by
\r{417})
\be
&&X_{\al}\bx^{\mu}=0,\quad X_{\al}\bp^{\mu}=0
\e{425}
as required by \r{319}.   The solutions
\r{409} are therefore of the exponential form
\r{3113}. The solutions \r{411} satisfy \r{413} since \r{334} is satisfied.

For the solutions \r{411} we find the first order coefficients
\be
&&X^{\mu}_1=0,\quad
X^{\mu}_2=-{x^{\mu}\over 2x^2},\quad P^{\mu}_1=-{p^{\mu}\over2(p\cdot x)},\quad
P^{\mu}_2={p^{\mu}\over 2x^2}-{p^2\over 2x^2(p\cdot x)}x^{\mu}.
\e{426}
The general expression \r{416} yields here
\be
&&f_1=0,\quad f_2={(p\cdot x)\over 2x^2},
\e{427}
which also satisfies the condition \r{334}. One may easily check that this
choice
make $K_{\al\beta}$ in \r{327} zero, which means that the vector fields \r{417}
commutes. Even the solutions \r{411} satisfy the properties \r{425} which means
that also they are of the exponential form \r{3113}.
 Note that $\{\theta^1, f_1\}=1$ for the solution \r{409} and
$\{\theta^2, f_2\}=1$ for the solution \r{411}. These properties are
probably not
accidental as will be explained in section 8.

\setcounter{equation}{0}
\section{Extended observables in theories with first class constraints}
Consider a dynamical system defined on a $2n$-dimensional symplectic
manifold $\cM$.
Let
\be
&&\phi_a(x)=0, \quad a=1,\ldots,m<n
\e{501}
be first class constraints, \ie let $\phi_a(x)$ satisfy the Poisson algebra
\be
&&\{\phi_a(x), \phi_a(x) \}={f_{ab}}^c(x)\phi_c(x),
\e{502}
where ${f_{ab}}^c(x)$ are structure functions. Even here we may define extended
observables along the lines of section 3. For this we need $m$ functions
$\chi^a(x)$,
$a=1,\ldots,m$,  satisfying the properties
\be
&&\{\chi^a(x), \chi^b(x)\}=0,\quad \left.\det\{\chi^a(x),
\phi_b(x)\}\right|_{\phi=0}\neq0.
\e{503}
Since $\chi^a(x)$ and $\phi_b(x)$ together act like the second class constraint
variables
$\theta^{\al}$ in section 3, the previous analysis applies. Thus, we may
define a
Dirac bracket where both $\chi^a(x)$ and $\phi_b(x)$ are Casimir functions.
By means
of the formula \r{217} we may define the Dirac bracket by
\be
&&\{f, g\}_D=\{f, g\}+{M^a}_bV^bfV_ag-{M^a}_bV_afV^bg-{f_{ab}}^c\phi_cV^afV^bg,
\e{504}
where
\be
&&{M^a}_b(x)\equiv\{\chi^a(x), \phi_b(x)\},
\e{505}
and
\be
&&V^a={(M^{-1})^a}_b(x)\{\chi^b,\cdot\},\quad
{(M^{-1})^a}_b(x){M^b}_c(x)=\del^a_c,\nn\\
&&V_a=-{(M^{-1})^b}_a(x)\{\phi_b,\cdot\}+
{(M^{-1})^c}_a(x){(M^{-1})^d}_b(x){f_{cd}}^e\phi_e\{\chi^b,\cdot\},
\e{506}
which are defined  according to formula \r{211}, \ie they are the
$V_\al$-operators
of section 3 for $\theta^\al=(\chi^a,\:\phi_a)$, and they differentiate the
Dirac
bracket \r{504}.

Let $x^{*i}$ be any solution of $\phi_a(x^*)=0$ where $m$ of the
coordinates $x^i$
are made dependent variables. $x^{*i}$ are viewed as observables here. By
extended
observables $\tx^i(x)$ we then mean expressions defined on $\cM$ satisfying the
properties
\be
&&\tx^i(x^*)=x^{*i},
\e{507}
\be
&&\phi_a(\tx)=0,
\e{508}
and
\be
&&\{\tx^i(x), \tx^j(x)\}=\{x^i,
x^j\}_D|_{x\ra\tx(x)}.
\e{509}
The first condition \r{507} requires $\tx^i(x)$ to be of the form
\be
&&\tx^i(x)=x^i+\sum_{n=1}^\infty
X^{ia_1\cdots a_n}(x)\phi_{a_1}(x)\cdots\phi_{a_n}(x).
\e{510}
It is  easily seen that
\be
&&\left.\tx^i(x)=e^{-\xi_aV^a}x^i\right|_{\xi\ra\phi}
\e{511}
are solutions of \r{508} of the form \r{510}. As in section 3 we have also the
property
\be
&&\left.\{x^i,
x^j\}_D|_{x\ra\tx(x)}= e^{-\xi_aV^a}\{x^i, x^j\}_D\right|_{\xi\ra\phi}=\{ \tx^i,
\tx^j\}_D,
\e{512}
where the last equality follows since $V^a$ differentiate the Dirac
bracket. Now since
the
$\chi^a$-variables are chosen to satisfy \r{503} the $V^a$-operators
commute, \ie
\be
&&[V^a, V^b]=0.
\e{513}
This implies that
\be
&&V^a\tx^i(x)=0\;\;\Leftrightarrow\;\;\{\chi^a, \tx^i(x)\}=0,
\e{514}
which for the Dirac bracket \r{504}   implies
\be
&&\{\tx^i,
\tx^j\}_D=\{ \tx^i,
\tx^j\}.
\e{515}
This together with \r{512} shows that the condition \r{509} is satisfied.
Notice that
in distinction to the extended observables $\bx^i(x)$ in the second class case,
$\tx^i(x)$ are much more ambiguous due to the large freedom how to choose the
``gauge fixing" variables $\chi^a$. In other words there is a large  gauge
freedom in $\tx^i(x)$.

\subsection{Example: The free relativistic particle}
Let $x^{\mu}$ and $p_{\mu}$ be coordinates and momenta for a free relativistic
particle. The momenta satisfy then the mass shell condition (we use
timelike metric)
\be
&&\phi\equiv p^2-m^2=0.
\e{516}
Let us choose as ``gauge fixing" variable
\be
&&\chi\equiv \eta\cdot x-\tau,
\e{517}
where $\tau$ is a parameter and $\eta^{\mu}$ a constant four-vector. The
condition
\r{503} requires $\eta^2\geq 0$. The extended observables are here
\be
&&\left.\tx^{\mu}=e^{-\xi V}x^{\mu}\right|_{\xi\ra\phi}, \quad
\left.\tpe_{\mu}=e^{-\xi
V}p_{\mu}\right|_{\xi\ra\phi}, \quad V\equiv {1\over2\eta\cdot p}\{\eta\cdot x,
\:\cdot\:\}.
\e{518}
Explicitly we have
\be
&&\tx^{\mu}=x^{\mu},\quad \tpe^{\mu}=p^{\mu}-{\eta^{\mu}\eta\cdot
p\over\eta^2}\left(1-\sqrt{1-{\eta^2(p^2-m^2)\over(\eta\cdot p)^2}}\right),\quad
\eta^2>0.
\e{519}
\be
&&\tx^{\mu}=x^{\mu},\quad \tpe^{\mu}=p^{\mu}-{\eta^{\mu}\over2\eta\cdot
p}(p^2-m^2),\quad
\eta^2=0.
\e{520}
If the particle is massive, $m\neq0$, we may  also choose the proper time
gauge fixing
\be
&&\chi\equiv p\cdot x-\tau.
\e{521}
In this case we have
the expressions \r{518} for
\be
&&V={1\over 2p^2}\{p\cdot x, \:\cdot\:\},
\e{522}
which explicitly yield
\be
&&\tx^{\mu}=\sqrt{{p^2\over m^2}}\,x^{\mu},\quad \tpe^{\mu}=\sqrt{{m^2\over
p^2}}\,p^{\mu}.
\e{523}
All solutions \r{519}, \r{520}, and \r{523} satisfy $\tpe^2=m^2$. One may easily
check that property \r{509} is valid.

\setcounter{equation}{0}
\section{Comparisons with gauge invariant extensions}
The extended observables constructed in the previous section are not what
one usually
considers to be observables in a gauge theory. Normally they are gauge invariant
quantities. Indeed there is something called
gauge invariant extensions in the literature (see \eg \cite{str,pol}). They are
quantities
$\hx^i(x)$ defined on $\cM$ with the following properties: They are gauge
invariant,
\ie they satisfy
\be
&&\{\hx^i(x), \phi_a(x)\}=C^{i\:b}_a(x)\phi_b(x),
\e{601}
where $\phi_a(x)$ are the first class constraint variables satisfying the
algebra
\r{502}, and $C^{i\:b}_a(x)$ are unspecified coefficient functions.
$\hx^i(x)$ also
satisfy
\be
&&\chi^a(\hx)=0,
\e{602}
where $\chi^a$ are  ``gauge fixing" variables satisfying the properties
\r{503}, and
\be
&&\hx^i(x^*)=x^{*i},
\e{603}
where $x^{*i}$ is any solution of $\chi^a(x^*)=0$ with $m$ dependent
coordinates.
Furthermore, we have
\be
&&\{\hx^i(x), \hx^j(x)\}=\{x^i,
x^j\}_D|_{x\ra\hx(x)}+C^{ij\:b}(x)\phi_b(x),
\e{604}
where $C^{ij\:b}(x)$ are unspecified coefficient functions and where the Dirac
bracket is the one in \r{504}. For Lie group theories the following general
formula
was given in
\cite{pol}
\be
&&\hx^i(x)=\int d^m\Omega\,
|\det\{\chi^a(x_{\Omega}),
\phi_b(x_{\Omega})\}|\,\del^m(\chi(x_{\Omega}))x^i_{\Omega},
\e{605}
where $x^i_{\Omega}$ is a gauge transformed $x^i$ and $d^m\Omega$ is the volume
element of the group.

>From the analysis of the previous sections we are now able to make a more
careful
analysis  of these gauge invariant  extensions. From the property \r{603}
it is clear
that
$\hx^i(x)$ must be of the general form
\be
&&\hx^i(x)=x^i+\sum_{n=1}^\infty X^i_{a_1\cdots
a_n}(x)\chi^{a_1}(x)\cdots\chi^{a_n}(x).
\e{606}
In fact, the obvious solution of this form is
\be
&&\left.\hx^i(x)= e^{-\xi^aV_a}x^i\right|_{\xi\ra\chi},
\e{607}
where the differential operator $V_a$ is given in \r{506}. This expression of
$\hx^i(x)$ satisfies \r{602} due to the properties
\be
&&V_a\chi^b=\del_a^b,\quad V_a\phi_b=0.
\e{608}
The properties \r{601} and \r{604} follow then from the following commutation
relations
\be
&&[V_a, V_b]=\phi_e(x) \{f_{cd}^{\;\;
e}(x)(M^{-1})^c_{\;a}(x)(M^{-1})^d_{\;b}(x),\:\cdot\:\}_D,
\e{609}
where the Dirac bracket is the one in \r{504}. Thus, the $V_a$-operators
only commute
if the structure functions $f_{ab}^{\;\;c}(x)$ and $M^a_{\;b}(x)$ are
functions of
only the constraint variables $\phi_a$ and/or $\chi^b$. In this case the last
terms in \r{604} vanish. Strict gauge invariance, \ie
\be
&&\{\hx^i(x), \phi_a(x)\}=0
\e{610}
is only valid for abelian gauge theories.

\subsection{Example: The free relativistic particle}
Let as in the previous section $x^{\mu}$ and $p_{\mu}$ be coordinates and
momenta for
a free relativistic particle where the momenta satisfy
\be
&&\phi\equiv p^2-m^2=0.
\e{611}
The gauge invariant extensions are easily calculated by means of formula
\r{605}. In
the gauge $\chi=\eta\cdot x-\tau$ we find
\be
&&\hp^{\mu}=p^{\mu},\quad \hx^{\mu}={\eta_{\nu}J^{\mu\nu}\over\eta\cdot
p}+\tau{p^{\mu}\over\eta\cdot p},\quad J^{\mu\nu}\equiv
x^{\mu}p^{\nu}-x^{\nu}p^{\mu},
\e{612}
and in the proper time gauge $\chi=x\cdot p-\tau$ we have
\be
&&\hp^{\mu}=p^{\mu},\quad \hx^{\mu}={p_{\nu}J^{\mu\nu}\over
p^2}+\tau{p^{\mu}\over
p^2}.
\e{613}
These expressions satisfy \r{602}-\r{603} and property \r{601} with $C^{i
b}_a=0$ and
\r{604} with $C^{ijb}=0$.

\setcounter{equation}{0}
\section{Extended observables in general gauge theories in a particular gauge}
General gauge theories are theories with first class constraints. In the
previous
sections, 5 and 6, we have constructed two types of extended observables
for these
theories. Here we define a third type namely extended observables as defined in
section 3 for second class constraints. Consider therefore again the first class
constraint variables
$\phi_a(x)$ in
\r{501}-\r{502} and the gauge fixing variables $\chi^a(x)$ with the properties
\r{503}.  The physical degrees of freedom are  described by
$x^{*i}$ satisfying the conditions
\be
&&\phi_a(x^*)=0,\quad\chi^a(x^*)=0,\quad a=1,\ldots,m.
\e{701}
$x^{*i}$ depends on  $2(n-m)$ independent coordinates.
As in section 3 we view $x^{*i}$ as observables here. In order to define
extended
observables $\bx^i(x)$ we need  differential operators $X_{\al}$,
$\al=1,\ldots,2m$, which probably must be commuting. The
$V_{\al}$~-~operators in
\r{506} do not commute in general. However, we expect that we  always may define
commuting
$X_{\al}$-operators defined by
\be
&&X^a\equiv -V^a-\{f^a,\:\cdot\:\}_D,\nn\\
&&X_a\equiv
-V_a-\phi_e\{f^e_{\:a},\:\cdot\:\}_D,
\e{702}
where $V^a$ and $V_a$ are the $V_{\al}$-operators in \r{506}, and where
$f^a$ and
$f^e_{\:a}$ satisfy the condition
\be
&&\{f^c_{\:a}, \bx^i\}_DM^a_{\:b}\{f^b, \bx^j\}_D-\{f^c_{\:a},
\bx^j\}_DM^a_{\:b}\{f^b,
\bx^i\}_D+\{f^a, \bx^i\}_Df_{ab}^{\;\;c}\{f^b, \bx^j\}_D=0,\nn\\
\e{7021}
which follows from \r{334}.
 The extended observables are then given by
\be
&&\left.\bx^i(x)=e^{\xi_aX^a+\rho^aX_a}x^i\right|_{\xi\ra\phi, \rho\ra\chi}.
\e{703}
Even these expressions are gauge invariant in the sense of \r{601}. They may
therefore be called proper gauge invariant extensions since they describe
exactly the
right number of degrees of freedom in distinctions to the gauge invariant
extensions
in section 6.

\subsection{Example: The free relativistic particle}
Consider again the relativistic particle with coordinates
 $x^{\mu}$ and momenta $p_{\mu}$  satisfying the mass shell condition
\be
&&\phi\equiv p^2-m^2=0.
\e{704}
In the following we denote the $V_{\al}$-operators by $V$ and $W$
corresponding to
$V^a$ and
$V_a$ respectively. In the proper time gauge
\be
&&\chi=p\cdot x-\tau,
\e{705}
we have then
\be
&&V\equiv{1\over2p^2}\{p\cdot x,\:\cdot \:\}\quad
W\equiv-{1\over2p^2}\{p^2,\:\cdot
\:\},
\e{706}
which do commute. The extended observables are therefore
\be
&&\left.\bx^{\mu}(x,p)=e^{-\xi V-\rho W}x^{\mu}\right|_{\xi\ra\phi,
\rho\ra\chi}=\tx^{\mu}(\hx, \hp)=\hx^{\mu}(\tx,
\tpe)=\sqrt{{p^2\over m^2}}\left({p_{\nu}J^{\mu\nu}\over p^2}+\tau{p^{\mu}\over
p^2}     \right),\nn\\
&&\left.\bp^{\mu}(x,p)=e^{-\xi V-\rho
W}p^{\mu}\right|_{\xi\ra\phi,
\rho\ra\chi}=\tpe^{\mu}(\hx, \hp)=\hp^{\mu}(\tx,
\tpe)=\sqrt{{m^2\over p^2}}p^{\mu},
\e{707}
where $\tx^{\mu}$ and $\tpe^{\mu}$ are given in \r{523}, and $\hx^{\mu}$,
$\hp^{\mu}$  in \r{613}.

 In the gauge
\be
&&\chi=\eta\cdot x-\tau,\quad \eta^2\geq0
\e{709}
we have
\be
&&V\equiv{1\over2\eta\cdot p}\{\eta\cdot x,\:\cdot \:\}\quad
W\equiv-{1\over2\eta\cdot p}\{p^2,\:\cdot
\:\}.
\e{710}
These operators do {\em not} commute. We have
\be
&&[V, W]=\{{1\over2\eta\cdot p},\:\cdot\:\}_D.
\e{711}
However, if we replace $V$ by $X$ defined by
\be
&&X\equiv V+\{{\eta\cdot x\over2\eta\cdot p},\:\cdot\:\}_D,
\e{712}
then $X$ and $W$ commute. This choice  satisfies the condition \r{7021}.
The extended observables are therefore for $\eta^2>0$
\be
&&\left.\bx^{\mu}(x,p)=e^{-\xi X-\rho W}x^{\mu}\right|_{\xi\ra\phi,
\rho\ra\chi}=\tx^{\mu}(\hx, \hp)=\hx^{\mu}(\tx,
\tpe)=\nn\\&&=\hx^{\mu}+{\tau\over\eta\cdot
p}\left({1-\sqrt{A}\over\eta^2\sqrt{A}}\right)\left(\eta^2 p^{\mu}-\eta\cdot p
\eta^{\mu}\right),\quad A\equiv 1-{\eta^2(p^2-m^2)\over(\eta\cdot p)^2},   \nn\\
&&\left.\bp^{\mu}(x,p)=e^{-\xi X-\rho W}p^{\mu}\right|_{\xi\ra\phi,
\rho\ra\chi}=\tpe^{\mu}(\hx, \hp)=\hp^{\mu}(\tx,
\tpe)=\tpe^{\mu},
\e{713}
where $\hx^{\mu}$ and $\hp^{\mu}$ are given in \r{612},  and where
$\tx^{\mu}$ and
$\tpe^{\mu}$ are given by
\be
&&\left.\tx^{\mu}(x)=e^{-\xi
X}x^{\mu}\right|_{\xi\ra\phi}=x^{\mu}+{\eta\cdot
x\over\eta\cdot p}\left({1-\sqrt{A}\over\eta^2\sqrt{A}}\right)
\left(\eta^2 p^{\mu}-\eta\cdot p
\eta^{\mu}\right),\nn\\&&\left.\tpe^{\mu}(x)=e^{-\xi
X}p^{\mu}\right|_{\xi\ra\phi}
=p^{\mu}-{\eta^{\mu}\eta\cdot
p\over\eta^2}\left(1-\sqrt{A}\right).
\e{714}
For $\eta^2=0$ these expressions reduce to
\be
&&\bx^{\mu}(x,p)=\hx^{\mu}-\tau{\eta^{\mu}\over2(\eta\cdot p)^2}(p^2-m^2),\nn\\
 &&\bp^{\mu}(x,p)=p^{\mu}-{\eta^{\mu}\over2\eta\cdot p}(p^2-m^2),
\e{715}
and
\be
&&\tx^{\mu}(x)=x^{\mu}-\eta^{\mu}{\eta\cdot
x\over2(\eta\cdot
p)^2}(p^2-m^2),\nn\\&&\tpe^{\mu}(x)=p^{\mu}-{\eta^{\mu}\over2\eta\cdot
p}(p^2-m^2).
\e{716}

\setcounter{equation}{0}
\section{Quantization}
We believe  the existence of extended observables to be a strong indication
that the
given theory  may be quantized in a covariant fashion. Since we know how
theories with
first class constraints may be quantized covariantly,  our main interest is
in theories
with second class constraints. The question then is  how the extended
observables and
their construction could be helpful in a covariant quantum theory. The
exponential
mapping to the extended observables could perhaps be used for the
construction of
physical symbols from the original ones (cf \cite{BO}). However, before we
give any
prescription for a covariant quantization of second class theories we need to
understand the properties obtained so far.  First it is clear that although
we have
succeeded to specify the properties of extended observables in details in
section 3,
we have not yet obtained a precise classification of all theories for which
these
observables actually exist. This will probably be clarified in the near
future since
we have obtained simple general forms of the solutions and simple conditions for
their existence. From the very simple examples treated in section 4 and
comparisons
with corresponding properties for theories with first class constraints, it is
obvious that the approach here is directly connected to the approach of
splitting the
second class constraints  into first class ones and gauge fixing conditions
\cite{FS,HM}. In principle such a splitting is always possible (a
polarization). One
may notice that for the simple but nontrivial example treated in subsection
4.2 we
found two distinct extended observables. From the point of view of sections
5-7 it is
clear that these two solutions correspond to two natural choices of gauge
generators.
In fact, $\theta^1$ is a gauge generator for the solution \r{409} and
$\theta^2$ for \r{411}. The extended observables are therefore gauge
invariant from
this point of view and can also be viewed as  proper gauge invariant
extensions of
the type given in section 7. The example in 4.2 seems also to give a clue for
what the $f_{\al}$-functions actually do for us. The commuting
$X_{\al}$-operators for the solution \r{409} has the property
$X_{\al}A=0\;\Leftrightarrow\;\{f_1, A\}=\{\theta^1, A\}=0$ which is
consistent since
$\{\theta^1, f_1\}=1$. (For the solution
\r{411} we have $X_{\al}A=0\;\Leftrightarrow\;\{f_2, A\}=\{\theta^2, A\}=0$ and
$\{\theta^2, f_2\}=1$.) The same happens in the example of section 7. Thus,
it seems
as if the nonzero
$f_{\al}$-functions replace an equal number of $\theta^{\al}$-functions in
such a
fashion that $X_{\al}$ will commute. These observations should play an
important role
in a covariant quantization.

One way to quantize theories with second class constraints covariantly in
which the
extended observables play a crucial role may be described as follows: Consider a
theory with a Hamiltonian $H(x)$ and first and second class constraints
$\phi_a$ and
$\theta^{\al}$. Eliminate the second class constraints $\theta^{\al}$ by
replacing
$x^i$ by  the extended observables $\bx^i(x)$ in $H(x)$ and $\phi_a(x)$ where
$\bx^i(x)$ is constructed according to section 3. The theory is then
transformed into
an equivalent theory given by the Hamiltonian
$H(\bx)$ and the first class constraints
$\phi_a(\bx)$ and the  gauge generators under which $\bx^i(x)$ is gauge
invariant. \\

\noindent
{\bf Example: Particle on a sphere.}

Consider a theory with the Hamiltonian (a free nonrelativistic particle)
\be
&&H={\bfp^2\over2m},
\e{801}
and the constraints $\theta^{\al}=0$ where
\be
&&\theta^1=\bfx^2-R^2, \quad  \theta^2=\bfp\cdot\bfx,
\e{802}
where  $R$ is a positive constant (cf subsection 4.1). The resulting theory
describes a free particle on a sphere with radius $R$. The extended
observables are
here
\be
&&\bar{\bfx}={R\over\sqrt{\bfx^2}}\bfx, \quad \bar{\bfp}={\sqrt{\bfx^2}\over
R}\left(\bfp-{\bfp\cdot\bfx\over\bfx^2}\bfx\right).
\e{803}
 We may therefore eliminate $\theta^{\al}$ and consider the
equivalent theory with the Hamiltonian
\be
&&H={\bar{\bfp}^2\over2m}={{\bf L}^2\over2mR^2}, \quad {\bf L}=\bfx\times\bfp
\e{804}
Quantization yields then the spectrum
\be
&&E_l={\hbar^2 l(l+1)\over2mR^2}.
\e{805}
The state space is restricted by either $\theta^{1}$ or $\theta^{2}$ as gauge
generator. This procedure should be compared to the  treatment of \cite{HM}
in which
the method of splitting the second class constraints in gauge generators
and gauge
fixings is used. The difference is that we here are making use of uniquely
defined
extended observables which  precisely define the equivalent theory.

\setcounter{equation}{0}
\section{Summary of the results}
In this paper we have obtained new results for general classical
Hamiltonian theories
with constraints. We have for the first time precisely defined and explicitly
constructed extended observables for such theories which considerably
generalizes the
concept of gauge invariant extensions used in general gauge theories. (Even the
properties of the latter are further clarified.) For simplicity we have
considered
finite dimensional theories. (The generalization to infinite dimensional
theories is
in principle straight-forward.) The results may be summarized as follows:

Given is a $2n$-dimensional symplectic manifold $\cM$ with coordinates $x^i$,
$i=1,\ldots,2n$. Its closed,  non-degenerate two-form and related Poisson
bracket is
defined in section 2. On this manifold we have a set of constraints. They
may either
be of first or of second class (or a mixture) in Dirac's classification
\cite{Dirac}.
Consider first the case of second class constraints. We have then the constraints
$\theta^{\al}(x)=0$, $\al=1,\ldots,2m<2n$, satisfying the properties
\be
&&\left.{\det C^{\al\beta}}\right|_{\theta=0}\neq0, \quad
C^{\al\beta}\equiv\{\theta^{\al},
\theta^{\beta}\}.
\e{901}
In this case the extended observables $\bx^i(x)$ are defined as follows:
$\bx^i(x)$
are functions on $\cM$ satisfying the properties
\be
1)&& \bx^i(x^*)=x^{*i}
\e{902}
for any solution $x^{*i}$ of $\theta^{\al}(x^*)=0$. ($x^{*i}$ are
observables.) Also
the extended observables themselves are solutions of the constraints,
\be
2)&& \theta^{\al}(\bx(x))=0,\quad \al=1,\ldots,2m.
\e{903}
Furthermore, they satisfy the Poisson algebra
\be
3)&& \{\bx^i(x), \bx^j(x)\}=\{x^i,
x^j\}_D|_{x\ra\bx(x)},
\e{904}
where the bracket on the left-hand side is the original Poisson bracket  on
$\cM$, while the bracket on the right-hand side is the Dirac bracket defined in
\r{205}. The general solutions of conditions 1)-3) we have found to be of the
following form
\be
&&\bx^i(x)=\left. e^{\xi^{\al}X_{\al}}x^i\right|_{\xi=\theta},
\e{905}
where $\xi^{\al}$ are parameters and $X_{\al}$ vector fields of the form
\be
&&X_{\al}=-V_{\al}+\{f_{\al},\:\cdot\:\}_D,\quad
V_{\al}=C_{\al\beta}\{\theta^{\beta},\:\cdot\:\}.
\e{906}
$X_{\al}$ differentiates the Dirac bracket \r{205} since $V_{\al}$ also has this
property as proved in the appendix. The functions $f_{\al}$ in \r{906} must
be chosen
such that
\be
&&V_{\al}\bx^i(x)C^{\al\beta}V_{\beta}\bx^j(x)=0
\e{907}
 is satisfied. We believe that this requires $f_{\al}$ to be chosen such that
$X_{\al}$ in \r{906} commute and at the same time satisfy
\be
&&\{f_{\al},\bx^i\}_DC^{\al\beta}\{f_{\beta},\bx^j\}_D=0.
\e{908}
However, the last two conditions seem to be stronger than \r{907}. In order for
$X_{\al}$ to commute
$f_{\al}$ must be chosen such that $K_{\al\beta}$ defined by
\be
&&K_{\al\beta}=C_{\al\beta}+\{f_{\al}, f_{\beta}\}_D+C_{\beta\ga}\{\theta^{\ga},
f_{\al }\}-C_{\al\ga}
\{\theta^{\ga}, f_{\beta}\}
\e{909}
 at most depends on the constraint variables $\theta^{\al}$ apart from
constants (see
\r{326}).

For first class constraints $\phi_a(x)=0$, $a=1,\ldots,m<n$ we may define extended
observables $\tx^i(x)$ analogously. $\tx^i(x)$ satisfies
\be
1)&& \tx^i(x^*)=x^{*i}
\e{910}
for any solution $x^{*i}$ of $\phi_a(x^*)=0$, and $\tx^i(x)$ are themselves
solutions,
\be
2)&&\quad \phi_a(\tx^i(x))=0,\quad a=1,\ldots,m.
\e{911}
Furthermore, we have the relation
\be
3)&& \{\tx^i(x), \tx^j(x)\}=\{x^i,
x^j\}_D|_{x\ra\tx(x)},
\e{912}
where the Dirac bracket is defined in terms of $\phi_a$ and a set of gauge
fixing
variables $\chi^a(x)$ satisfying
\be
&&\{\chi^a(x), \chi^b(x)\}=0,\quad \left.\det\{\chi^a(x),
\phi_b(x)\}\right|_{\phi=0}\neq0.
\e{913}
The solution is
\be
&&\left.\tx^i(x)=e^{-\xi_aV^a}x^i\right|_{\xi\ra\phi},
\e{914}
where
\be
&&V^a={(M^{-1})^a}_b(x)\{\chi^b,\cdot\},\quad {M^a}_b(x)\equiv\{\chi^a(x),
\phi_b(x)\}.
\e{915}
These $V^a$-operators commute due to \r{913}. The extended observables
\r{914} are not
what one normally would call observables in a general gauge theory since
they are not
gauge invariant. In section 6 we defined gauge invariant observables
$\hx^i(x)$ along
the lines what has been considered before (see \eg \cite{str,pol}). They satisfy
$\chi^a(\hx)=0$ instead of 1) in \r{910} and a weak form of 3) in \r{912}. If we
actually make use of $\chi^a$ as gauge fixing, \ie view $\phi_a(x)=0$ and
$\chi^a(x)=0$ as second class constraints then we may construct the
corresponding
extended observables according to the rules for second class constraints.
We find
then the solution \r{905} with $\theta^{\al}=(\phi_a, \chi^a)$.
Interestingly enough
this solution is also gauge invariant and  it could be called the proper gauge
invariant extension. The method to construct extended observables for
second class
constraints seems therefore to be appropriate also for first class constraints.

In section 8 we discussed the possibility to perform a covariant quantization of
theories with second class constraints using the new insights of the
present paper.
First we noticed that the results so far seem to indicate a connection to
the method
of splitting the second class constraints into first class ones and gauge fixing
conditions (see \cite{FS,HM}). However, the extended observables by
themselves also
provide for a simple algorithm how to map the constrained theory to an
equivalent
unconstrained one. This mapping procedure was exemplified for a particle on
a sphere
and led to a very simple quantization of this system. Whether or not such a
procedure
eventually may compete with the standard conversion method \cite{BF} in
which the
second class constraints are converted to first class ones by means of
additional
variables are left for the future to decide.

\vspace{0.5cm}

{\bf Acknowledgments}: One of the authors (S.L.) thanks Lars Brink for his warm hospitality at the institute of theoretical physics, Chalmers and G\"oteborg university.
This work is partially supported by INTAS grant 00-262.
SL acknowledges support from Russian Ministry of Education under the
grant E-00-33-184 and from Russian Foundation for Basic Research grant no
00-02-17-956.

\begin{appendix}
\setcounter{equation}{0}
\section{Proof of Leibniz' rule (2.14) for $V_{\al}$}
Let us prove \r{213} backwards:
\be
&&\{V_{\al}f, g\}_D+\{f, V_{\al}g\}_D=\{\{f, \theta^{\beta}\}C_{\beta\al},
g\}_D+\{f,
\{g, \theta^{\beta}\}C_{\beta\al}\}_D=\nn\\&&=\{\{f,
\theta^{\beta}\}C_{\beta\al},
g\}-\{\{f, \theta^{\beta}\}C_{\beta\al}, \theta^{\ga}\}C_{\ga\la}\{\theta^{\la},
g\}+\nn\\&&+\{f, \{g, \theta^{\beta}\}C_{\beta\al}\}-\{f,
\theta^{\ga}\}C_{\ga\la}\{\theta^{\la}, \{g,
\theta^{\beta}\}C_{\beta\al}\}=\nn\\&&=
\left(\{\{f, \theta^{\beta}\}, g\}+\{f, \{g,
\theta^{\beta}\}\}\right)C_{\beta\al}+\nn\\&&+
\{f,\theta^{\ga}\}\{C_{\ga\al},g\}+\{ C_{\la\al}, f\}\{\theta^{\la},
g\}-\nn\\&&-
\{\{f, \theta^{\beta}\}, \theta^{\ga}\}C_{\beta\al}C_{\ga\la}\{\theta^{\la},
g\}-\{f, \theta^{\ga}\}C_{\ga\la}C_{\beta\al}\{\theta^{\la}, \{ g,
\theta^{\beta}\}\}-\nn\\&&-\{ f, \theta^{\ga}\}\{C_{\ga\al},
\theta^{\beta}\}C_{\beta\la}\{\theta^{\la}, g\}-\{ f,
\theta^{\ga}\}\{\theta^{\beta}, C_{\la\al}\}C_{\ga\beta}\{g, \theta^{\la}\}.
\e{a1}
Inserting
\be
&&\{C_{\ga\al}, g\}=-C_{\ga\la}\{C^{\la\beta},
g\}C_{\beta\al}=|Jac.\:id.|=\nn\\&&=C_{\ga\la}\{\{\theta^{\beta}, g\},
\theta^{\la}\}C_{\beta\al}+C_{\ga\la}\{\{g, \theta^{\la}\},
\theta^{\beta}\}C_{\beta\al},
\e{a2}
and the corresponding expression  for $\{C_{\la\al}, f\}$ into \r{a1} we find
\be
&&\{V_{\al}f, g\}_D+\{f, V_{\al}g\}_D=
\{f,  g\}C_{\beta\al}-
\{\{f, \theta^{\beta}\}, \theta^{\ga}\}C_{\beta\al}C_{\ga\la}\{\theta^{\la},
g\}-\nn\\&&-\{f, \theta^{\ga}\}C_{\ga\la}C_{\beta\al}\{\theta^{\la}, g\}
\theta^{\beta}\}\}-\{\{ f, \theta^{\ga}\},
\theta^{\beta}\}C_{\ga\la}\{\theta^{\la}, g\}C_{\beta\al}-\nn\\&&-
\{ f,
\theta^{\ga}\}\left(\{
C_{\ga\al}, \theta^{\beta}\}C_{\beta\la}+\{
C_{\al\la}, \theta^{\beta}\}C_{\beta\ga}\right)\{\theta^{\la}, g\},
\e{a3}
where we also have made use of a Jacobi identity for the first terms. Now
we have
\be
&&\{
C_{\ga\al}, \theta^{\beta}\}C_{\beta\la}+cycle(\ga, \al, \la)=0,
\e{a4}
which is easily proved by means of Jacobi identities. \r{a4} in \r{a3}
yields then
\be
&&\{V_{\al}f, g\}_D+\{f,
V_{\al}g\}_D=\nn\\&&=\{\{f,g\}-\{f,\theta^{\ga}\}C_{\ga\la}\{\theta^{\la}, g\},
\theta^{\beta}\}C_{\beta\al}=V_{\al}\{f, g\}_D
\e{a5}

\end{appendix}

\end{document}